\documentclass{article}
\usepackage{amssymb}

\usepackage{graphicx}
\usepackage{amsmath}
\usepackage{epstopdf}
\usepackage[utf8]{inputenc}
\usepackage[T1]{fontenc}

\begin{document}

\title{Quantum filtering equations for system driven
by non-classical fields}

\author{Anita D\c{a}browska\\ Nicolaus Copernicus University in Toru\'{n},
\\ Collegium Medicum in Bydgoszcz,\\ul. Jagiello\'{n}ska 15, 85--067 Bydgoszcz, Poland}

\date{}
\maketitle

\begin{abstract}
Using Gardiner and Collet's input-output model and the concept of cascade system, we determine the filtering equation for a quantum system driven by chosen non-classical states of light. The quantum system and electromagnetic field are described by making use of quantum stochastic unitary evolution.
We consider two examples of the non-classical states of the field: a combination of vacuum and single photon states and a mixture of two coherent states.
We describe the stochastic evolution conditioned on the results of the photon counting and quadrature measurements.
\end{abstract}


\abstract{Using Gardiner and Collet's input-output model and the concept of cascade system, we determine the filtering equation for a quantum system driven by chosen non-classical states of light. The quantum system and electromagnetic field are described by making use of quantum stochastic unitary evolution.
We consider two examples of the non-classical states of the field: a combination of vacuum and single photon states and a mixture of two coherent states.
We describe the stochastic evolution conditioned on the results of the photon counting and quadrature measurements.}



\section{Introduction}

Non-classical states of light such as continuous-mode Fock states \cite{L00}, their mixtures and superpositions arouse an interest because of their potential application in quantum metrology \cite{GLM11,LSZV12} and quantum communication \cite{KLM01,RRN05}. The theoretical description of the optimal excitation of quantum systems by wave packets travelling in space became one of the essential aspects of the subject \cite{SAL09,SAL10,WMSS11,E12,MS16}. To examine an excitation or a process of a storing of information in a single ion, atom or quantum dot master equations are usually considered. However, a detailed investigation of the phenomena of the excitation of quantum systems by wave packets travelling in space requires consideration of individual quantum trajectories conditional on the results of measurement of the light emitted or scattered by the system of interest. Derivation of master equations for a quantum system interacting with a wave packet of light prepared in a continuous-mode single photon state one can find, for instance, in \cite{GEPZ98,GJN11}, and for a wave packet taken in a continuous-mode Fock states in \cite{BCBC12}.

The main theme of this paper is determination of stochastic differential equations for quantum system interacting with chosen non-classical states of light. Let us make clear that we appeal to the theory of quantum state estimation \cite{Car93,BP02,Bar06,GZ10,WM10} based on the idea of the non-demolition observation \cite{B05} developed in the framework of the It\^{o} quantum stochastic calculus \cite{HP84,Par92}. So we consider an open quantum system interacting with an environment modelled by the Bose field. The Bose field, being an approximation of the electromagnetic field, disturbs the free evolution of the quantum system but it also allows for a continuous in time indirect observation of the system. An observation of the output field \cite{GarCol85} (the Bose field after interaction with the system) provides us with the information about the system. The stochastic equation describing the time evolution of the quantum system conditioned on the results of the continuous in time measurement performed on the output field is called quantum filtering equation. Solutions to this equation are called quantum trajectories \cite{Car93}. If the Bose field is prepared in the Gaussian state the evolution of the open system is Markovian and taking an average over all possible outcomes we get then from the stochastic equation the Gorini–-Kossakowski-–Sudarshan-–Lindblad master equation. Derivation of the filtering equation for the Bose field in the vacuum state one can find, for instance, in \cite{Car93,BB91,WM93,BGM04}. Determination of the conditional dynamics of the system interacting with the Bose field prepared in the squeezed Gaussian state was given in \cite{WM10}, and for multi-channels case in \cite{DG16}.

The evolution of open quantum system driven by non-classical state of light becomes non-Markovian and the standard methods of determination of the filtering equation do not work. Our article is motivated mainly by the papers \cite{GJN11,GJNC12} and it can be considered as a companion to them or as their successor. The mentioned papers treated the problem of determination of the filtering equations for a continuous-mode single photon state and continuous-mode cat states. We derive the filtering equation for the two cases: for the case when the Bose field is prepared in a combination of the vacuum and single photon states and for the case when the Bose field is in a mixture of two coherent cases.
In order to obtain the conditional evolution of the open system, we use the idea of introducing an ancilla being a source of the desired non-classical signal. We assume that the ancilla is driven by the vacuum and the output from ancilla is the input for the system.
We write down the filtering equations for the extended system consisting of the ancilla and the system of our interest and to get the conditional evolution of the system we simply take the partial trace of all elements of this equation with respect to ancilla. In this way we derive the stochastic equations for the conditional density matrix of the open system. We would like to stress that this equations were not determined in \cite{GJN11,GJNC12}. We give also the formulae for the probabilities of trajectories for the counting process not stated in \cite{GJN11,GJNC12}.

\section{Filtering equation for system driven by the field in a combination of vacuum and single photon states}

Let us introduce one-dimensional Bose field described by
\begin{equation}
B_{t}=\int_{0}^{t}b_{s}ds,\;\;\;B^{\dagger}_{t}=\int_{0}^{t}b^{\dagger}_{s}ds,\;\;\;
\Lambda_{t}=\int_{0}^{t}b^{\dagger}_{s}b_{s}ds,
\end{equation}
where the operators $b_{t}$, $b_{t}^{\dagger}$ satisfy the canonical commutation relations of the form
\begin{equation}
[b_{t},b_{s}]=0,\;\;\;[b_{t},b^{\dagger}_{s}]=\delta(t-s).
\end{equation}
The Hilbert space of the Bose field $\mathfrak{h}$ is the symmetric Fock space over $\mathbb{C}\otimes L^2(\mathbb{R}_{+})$. The space $\mathfrak{h}$ has a continuous tensor product structure, it means that it can be split it into the ``past and future spaces'' $\mathfrak{h}=\mathfrak{h}_{[0,t)}\otimes \mathfrak{h}_{[t,+\infty)}$, where $\mathfrak{h}_{[0,t)}$, $\mathfrak{h}_{[t,+\infty)}$ are the symmetric Fock spaces respectively over $\mathbb{C}\otimes L^2([0,t))$ and $\mathbb{C}\otimes L^2([t,+\infty))$ \cite{Bar06,HP84,Par92}.

The vacuum state in $\mathfrak{h}$ has a factorization property
\begin{equation}
|\mathit{vac}\rangle=|\mathit{vac}_{[0,t)}\rangle \otimes|\mathit{vac}_{[t,+\infty)}\rangle.
\end{equation}
The mean values of the increments
\begin{equation}
dB_{t}=B_{t+dt}-B_{t},\;\;\; dB^{\dagger}_{t}=B^{\dagger}_{t+dt}-B^{\dagger}_{t},\;\;\; d\Lambda_{t}=\Lambda_{t+dt}-\Lambda_{t}
\end{equation}
and their products for the vacuum state are
\begin{equation}\langle vac|dB_{t}|vac\rangle= 0,\;\; \langle vac|d\Lambda_{t}|vac\rangle =0,
\end{equation}
$$\langle vac|dB_{t}dB^{\dagger}_{t}|vac\rangle = dt,\;\;
\langle vac|dB^{\dagger}_{t}dB_{t}|vac\rangle = 0, \;\;\;
\langle vac|d\Lambda_{t}dB^{\dagger}_{t}|vac\rangle = 0.$$

We assume that the Bose field interacts with a quantum system (we will call it briefly $\mathcal{S}$) and the evolution of the composed system is given by the unitary operator, $U_{t}$, satisfying the It\^{o} quantum stochastic differential equation (QSDE) of the form \cite{Bar06,HP84,Par92}
\begin{eqnarray}
&&\mathrm{d}U_{t}=\left[L\mathrm{d}B^{\dagger}_{t}
- L^{\dagger}S\mathrm{d}B_{t}+\left(S-I\right)d\Lambda_{t}-
\left(\mathrm{i}
H_{\mathcal{S}}+\frac{1}{2}L^{\dagger}L
\right)\mathrm{d}t\right]U_{t},\nonumber\\
&&U_{t=0}=I,
\end{eqnarray}
where $L$, $S$, $H_{\mathcal{S}}$ are the operators acting in  $\mathfrak{h}_{\mathcal{S}}$ - the Hilbert space of the system $\mathcal{S}$, $H_{\mathcal{S}}$ stands for the Hamiltonian of $\mathcal{S}$, and $S$ is the unitary operator describing a process of a direct scattering of light by the system $\mathcal{S}$ \cite{Bar06}. Let us remind that the above equation is written in the interaction picture with respect to free dynamics of the Bose field. The unitary operator $U_{t}$ acts non-trivially only in $\mathfrak{h}_{\mathcal{S}}\otimes\mathfrak{h}_{[0,t)}$ and it commutes with the  increments $dB_{t}$, $dB^{\dagger}$, $d\Lambda_{t}$ acting non-trivially only in $\mathfrak{h}_{\mathcal{S}}\otimes\mathfrak{h}_{[t,t+dt)}$ \cite{Bar06}. According to the interpretation given by Gardiner and Collet, the operators $B_{t}$, $B^{\dagger}_{t}$, and $\Lambda_{t}$ describe the input Bose field -- the field before interaction with $\mathcal{S}$. And the output field -- the field after interaction with $\mathcal{S}$ -- is given by $B^{out}_{t}=U^{\dagger}_{t}B_{t}U_{t}$,
$B^{out\dagger}_{t}=U^{\dagger}_{t}B^{\dagger}_{t}U_{t}$, $\Lambda^{out\dagger}_{t}=U^{\dagger}_{t}\Lambda^{\dagger}_{t}U_{t}$ \cite{GZ10,GarCol85}.

\subsection{Continuous-mode single photon state}

The continuous-mode single photon state is defined as \cite{M08,M12}
\begin{equation}
|1_{\xi}\rangle= \int_{0}^{+\infty}\xi(t)dB^{\dagger}_{t}|\mathit{vac}\rangle
\end{equation}
with $\xi \in \mathbb{C}$ and the normalization
$\langle 1_{\xi}|1_{\xi}\rangle = \int_{0}^{\infty}|\xi(t)|^2dt = 1$.
In the frequency domain it has the form
\begin{equation}
|1_{\xi}\rangle= \int_{-\infty}^{+\infty} d\omega\widetilde{\xi}(\omega)b^{\dagger}(\omega)|\mathit{vac}\rangle,
\end{equation}
where $\widetilde{\xi}$ is the Fourier transform of $\xi$.

The mean values of the increments $dB_{t}$, $dB_{t}^{\dagger}$, $d\Lambda_{t}$ and their products for the Bose field in the single photon state are
$$\langle 1_{\xi}|dB_{t}|1_{\xi}\rangle =0,\;\;\; \langle 1_{\xi}|d\Lambda_{t}|1_{\xi}\rangle = |\xi(t)|^2dt,$$
\begin{equation}
\langle 1_{\xi}|dB_{t}dB^{\dagger}_{t}|1_{\xi}\rangle= dt,
\;\;\;
\langle 1_{\xi}|dB^{\dagger}_{t}dB_{t}|1_{\xi}\rangle= 0,\;\;\;
\langle 1_{\xi}|d\Lambda_{t}dB^{\dagger}_{t}|1_{\xi}\rangle = 0.
\end{equation}

Let us notice that for the single photon state we have the additive decomposition property
\begin{eqnarray}
|1_{\xi}\rangle =|1_{\xi[0,t)}\rangle\otimes |\mathit{vac}_{[t,+\infty)}\rangle
+|\mathit{vac}_{[0,t)}\rangle\otimes |1_{\xi[t,+\infty)}\rangle,
\end{eqnarray}
where
\begin{equation}
|1_{\xi[0,t)}\rangle\otimes |\mathit{vac}_{[t,+\infty)}\rangle=\int_{0}^{t}\xi(s)dB^{\dagger}_{s}|\mathit{vac}\rangle
\end{equation}
and
\begin{equation}
|\mathit{vac}_{[0,t)}\rangle\otimes |1_{\xi[t,+\infty)}\rangle
=\int_{t}^{+\infty}\xi(s)dB^{\dagger}_{s}|\mathit{vac}\rangle.
\end{equation}
By taking trace over the space $\mathfrak{h}_{[t,+\infty)}$, we get
\begin{eqnarray}
\mathrm{Tr}_{\mathfrak{h}_{[t,+\infty)}}|1_{\xi}\rangle\langle 1_{\xi}|&=& |1_{\xi[0,t)}\rangle \langle 1_{\xi[0,t)}|\nonumber\\&&+|\mathit{vac}_{[0,t)}\rangle \langle \mathit{vac}_{[0,t)}|\int_{t}^{+\infty}|\xi(s)|^2ds,
\end{eqnarray}
where $\int_{t}^{+\infty}|\xi(s)|^2ds$ is the probability that we do not measure the photon in the interval $[0,t)$. Moreover, for any bounded operator of the Bose field having the form $R_{t}=R_{[0,t)}\otimes I_{[t,+\infty)}$ (a bounded operator acting trivially on $\mathfrak{h}_{[t,+\infty)}$) we obtain
\begin{eqnarray}
\langle 1_{\xi}|R_{t}|1_{\xi}\rangle&=&\langle 1_{\xi[0,t)}|R_{[0,t)}|1_{\xi[0,t)}\rangle\nonumber\\
&&+ \langle \mathit{vac}_{[0,t)}|R_{[0,t)}|\mathit{vac}_{[0,t)}\rangle \int_{t}^{+\infty}|\xi(s)|^2ds.
\end{eqnarray}

\subsection{The reduced dynamics of $\mathcal{S}$}

Any bounded operator $X$ of the system $\mathcal{S}$ in the Heisenberg  picture,
\begin{equation}
j_{t}(X)=U_{t}^{\dagger}(X\otimes I)U_{t}\,,
\end{equation}
is an adapted process, which means that it acts as the identity in the space $\mathfrak{h}_{[t,+\infty)}$. To derive the differential equation for $j_{t}(X)$, we apply the rules of the quantum stochastic calculus of It\^{o} type (QSC) taking
\begin{equation}
dj_{t}(X)=dU_{t}^{\dagger}(X\otimes I)U_{t}+U_{t}^{\dagger}(X\otimes I)dU_{t}+dU_{t}^{\dagger}(X\otimes I)dU_{t}
\end{equation}
and using the table
\begin{eqnarray}
dB_{t}dB_{t}^{\dagger}=dt,\;\;\;dB_{t}d\Lambda_{t}=dB_{t},\nonumber\\
d\Lambda_{t}d\Lambda_{t}=d\Lambda_{t},\;\;\;d\Lambda_{t}dB_{t}^{\dagger}=dB_{t}^{\dagger}
\end{eqnarray}
with all the others products, including the products involving $dt$, vanishing.
In this way we obtain the stochastic differential equation of the form
\begin{eqnarray}
dj_{t}(X)&=&j_{t}\left(\mathcal{L}^{\ast} X\right)dt+j_{t}\left(S^{\dagger}[X,L]\right)dB^{\dagger}_{t}
\nonumber\\
&&+j_{t}\left([L^{\dagger},X]S\right)dB_{t}+j_{t}(S^{\dagger}XS-X)d\Lambda_{t}
\end{eqnarray}
with the superoperator
\begin{equation}
\mathcal{L}^{\ast}X= i[H, X]+L^{\dagger}X L-\frac{1}{2}L^{\dagger}LX-\frac{1}{2}X L^{\dagger} L\,.
\end{equation}

We assume that the compound system is prepared initially in the product state
\begin{equation}
\rho(0)\otimes\rho_{field}(0)\,.
\end{equation}
Thus the reduced density operator of $\mathcal{S}$ at the time $t$ is given by the formula
\begin{equation}
\rho(t)=\mathrm{Tr}_{\mathfrak{h}}\left(U_{t}\rho(0)
\otimes\rho_{field}(0)U_{t}^{\dagger}\right)\,.
\end{equation}
One can check, using the property
\begin{equation}
\mathrm{Tr}_{\mathfrak{h}_{S}}\left(X\rho(t)\right)=\mathrm{Tr}_{\mathfrak{h}_{\mathcal{S}}\otimes \mathfrak{h}}\left(j_{t}(X)\rho(0)
\otimes\rho_{field}(0)\right),
\end{equation}
that for the initial state of the Bose field of the form
\begin{eqnarray}\label{comb}
\rho_{field}(0)= \gamma_{00}|\mathit{vac}\rangle \langle \mathit{vac}|+\gamma_{01}|1_{\xi}\rangle \langle \mathit{vac}|+\gamma_{10}|\mathit{vac}\rangle \langle 1_{\xi}|+\gamma_{11}|1_{\xi}\rangle \langle 1_{\xi}|,
\end{eqnarray}
the reduced dynamics of $\mathcal{S}$ is given as
\begin{equation}
\rho(t)\;=\;\gamma_{00}\rho^{00}(t)+\gamma_{01}\rho^{10}(t)+\gamma_{10}\rho^{01}(t)
+\gamma_{11}\rho^{11}(t)\,,
\end{equation}
where the matrices $\rho^{00}(t)$, $\rho^{10}(t)$, $\rho^{01}(t)$, $\rho^{11}(t)$ satisfy the set of differential equations \cite{BCBC12}
\begin{eqnarray}\label{master1}
\dot{\rho}^{00}(t)&=&\mathcal{L}\rho^{00}(t),\nonumber\\
\dot{\rho}^{10}(t)&=&\mathcal{L}\rho^{10}(t)+[S\rho^{00}(t), L^{\dagger}]\xi(t),\nonumber\\
\dot{\rho}^{01}(t)&=&\mathcal{L}\rho^{01}(t)+[L,\rho^{00}(t)S^{\dagger}]\xi^{\ast}(t),\nonumber\\
\dot{\rho}^{11}(t)&=&\mathcal{L}\rho^{11}(t)+[S\rho^{01}(t),L^{\dagger}]\xi(t)+[L, \rho^{10}(t)S^{\dagger}]\xi^{\ast}(t)\nonumber\\
&&+\left(S\rho^{00}(t)S^{\dagger}-\rho^{00}(t)\right)|\xi(t)|^2,
\end{eqnarray}
where
\begin{equation}\label{superop}
\mathcal{L}\rho= -i[H, \rho]+L\rho L^{\dagger}-\frac{1}{2}L^{\dagger}L\rho-\frac{1}{2}\rho L^{\dagger}L\,.
\end{equation}
and initially $\rho^{00}(0)=\rho^{11}(0)= \rho(0)$, $\rho^{10}(0)=\rho^{01}(0)=0$. Note that the matrices $\rho^{10}(t)$ and $\rho^{01}(t)$ are non-Hermitian trace-class zero operators and  $\rho^{01}(t)=\left(\rho^{10}(t)\right)^{\dagger}$. Of course, the choice of the coefficients in the state (\ref{comb}) is constrained by the conditions that $\rho_{field}\geq 0$, $\rho_{field}=\rho_{field}^{\dagger}$, and $\mathrm{Tr}\{\rho_{field}\}=1$.

\subsection{The output processes}

Making use of QSC one can check the number operator of photons in the interval $t$ to $t+dt$ for the output field has the form
\begin{equation}
d\Lambda^{out}_{t}=d\Lambda_{t}+
j_{t}(L^{\dagger}S)dB_{t}+j_{t}(S^{\dagger}L)dB_{t}^{\dagger}+
j_{t}(L^{\dagger}L)dt\,.
\end{equation}
The mean value of $d\Lambda^{out}_{t}$ for the Bose field prepared in  (\ref{comb}) is
\begin{eqnarray}
\langle d\Lambda^{out}_{t}\rangle&=&\gamma_{11}|\xi(t)|^2dt+\mathrm{Tr}_{\mathfrak{h}_{\mathcal{S}}}
\left[L^{\dagger}S\left(\gamma_{11}\rho^{01}(t)+\gamma_{01}\rho^{00}(t)\right)\right]\xi(t)dt\nonumber\\
&&+\mathrm{Tr}_{\mathfrak{h}_{\mathcal{S}}}\left[S^{\dagger}L\left(\gamma_{11}\rho^{10}(t)
+\gamma_{10}\rho^{00}(t)\right)\right]\xi^{\ast}(t)dt\nonumber\\
&&+\mathrm{Tr}_{\mathfrak{h}_{\mathcal{S}}}\left(L^{\dagger}L\rho(t)\right)dt.
\end{eqnarray}

Using homodyne detection scheme we can measure the quadrature operator of the output field in the infinitesimal time increment from $t$ to $t+dt$,
\begin{equation}
dY(t)=dB^{out}_{t}+dB^{out\dagger}_{t}.
\end{equation}
One can check that
\begin{equation}
dB^{out}_{t}=j_{t}(S)dB_{t}+j_{t}(L)dt.
\end{equation}
So for the Bose field taken in (\ref{comb}) we have
\begin{eqnarray}
\langle dY(t)\rangle&=&
\mathrm{Tr}_{\mathfrak{h}_{\mathcal{S}}}[S(\gamma_{11}\rho^{01}(t)+\gamma_{01}\rho^{00}(t))]\xi(t)dt\nonumber\\
&&+\mathrm{Tr}_{\mathfrak{h}_{\mathcal{S}}}[S^{\dagger}(\gamma_{11}\rho^{10}(t)+\gamma_{10}\rho^{00}(t))]\xi^{\ast}(t)dt\nonumber\\
&&+\mathrm{Tr}_{\mathfrak{h}_{\mathcal{S}}}[(L+L^{\dagger})\rho(t)]dt.
\end{eqnarray}

\subsection{The generator of the Bose field in a combination of vacuum and single photon states}

As a generator of the Bose field in a combination of vacuum and single photon states, we consider a two level system which interacts with the Bose field in the vacuum state. We would like to stress that this system plays only a role of ancilla producing a signal in non-classical state. We assume that the evolution of the compound system (ancilla plus the Bose field) is given by the unitary operator, $\tilde{U_{t}}$, which satisfies QSDE
\begin{eqnarray}\label{uni1}
d\tilde{U_{t}}= \left(L_{A}dB_{t}^{\dagger}-L_{A}^{\dagger}dB_{t}
-\frac{1}{2}L_{A}^{\dagger}L_{A}dt\right)\tilde{U_{t}},\;\;\;\tilde{U}_{t=0}=I
\end{eqnarray}
with the coupling operator
\begin{equation}
L_{A}\;=\; \lambda(t)\sigma_{-}\,,
\end{equation}
where $\lambda(t)\in \mathbb{C}$ and $\sigma_{-}$ is the lowering operator from the excited $|1\rangle$ to the ground state $|0\rangle$ of ancilla. We set the Hamiltonian of ancilla $H_{A}=0$. The same generator was described in \cite{GJNC12}. The Schr\"{o}dinger equation for the state $|\Psi(t)\rangle=\tilde{U_{t}}|\psi_{0}\rangle\otimes |\mathit{vac}\rangle$ of the compound system then reads
\begin{equation}\label{schro}
d|\Psi(t)\rangle\;=\; \left(\lambda(t)\sigma_{-}dB_{t}^{\dagger}
-\frac{1}{2}\left|\lambda(t)\right|^{2}\sigma_{+}\sigma_{-}dt\right)|\Psi(t)\rangle\,.
\end{equation}
Let us notice that if the coupling coefficient
\begin{equation}\label{lambda}
\lambda(t)=\frac{\xi(t)}{\sqrt{\int_{t}^{+\infty}|\xi(s)|^2ds}}
\end{equation}
and if the initial state of the compound system is given as
\begin{equation}\label{inistate}
|\Psi(0)\rangle\;=\;\left(c_{0}|0\rangle+c_{1}|1\rangle\right)\otimes |\mathit{vac}\rangle
\end{equation}
then
\begin{eqnarray}
|\Psi(t)\rangle&=& c_{0}|0\rangle \otimes |\mathit{vac}\rangle+c_{1}
|1\rangle\otimes \sqrt{\int_{t}^{+\infty}|\xi(s)|^2ds}|\mathit{vac}\rangle\nonumber\\&&+c_{1}|0\rangle\otimes \int_{0}^{t}\xi(s)dB^{\dagger}_{s}|\mathit{vac}\rangle\,
\end{eqnarray}
is the exact solution of Eq. (\ref{schro}), and
\begin{equation}
\lim_{t\rightarrow +\infty} |\Psi(t)\rangle= |0 \rangle \otimes \left(c_{0}|\mathit{vac}\rangle +c_{1}|1_{\xi}\rangle\right).
\end{equation}
Thus, the Bose field after interaction with ancilla till time $t$ (the output field from ancilla) is in the state
\begin{eqnarray}
\rho_{field}^{out}(t)&=& \left|c_{0}\right|^2 |\mathit{vac}_{[0,t)}\rangle \langle \mathit{vac}_{[0,t)}|+ c_{0}c_{1}^{\ast} |\mathit{vac}_{[0,t)}\rangle \langle 1_{\xi[0,t)}|\nonumber\\
&&+c_{0}^{\ast}c_{1} | 1_{\xi[0,t)} \rangle\langle\mathit{vac}_{[0,t)}|+\left|c_{1}\right|^2 |1_{\xi[0,t)}\rangle \langle 1_{\xi[0,t)}|\nonumber\\&&+\left|c_{1}\right|^2 \int_{t}^{+\infty}|\xi(s)|^2ds |\mathit{vac}_{[0,t)}\rangle \langle \mathit{vac}_{[0,t)}|\,.
\end{eqnarray}
In particular, when the ancilla is prepared in the ground state, we simply get
\begin{eqnarray}
\rho_{field}^{out}(t)\;=\; |\mathit{vac}_{[0,t)}\rangle \langle \mathit{vac}_{[0,t)}|
\end{eqnarray}
and when ancilla is initially in the excited state, we have
\begin{eqnarray}
\rho_{field}^{out}(t)&=& |1_{\xi[0,t)}\rangle \langle 1_{\xi[0,t)}|+\int_{t}^{+\infty}|\xi(s)|^2ds |\mathit{vac}_{[0,t)}\rangle \langle \mathit{vac}_{[0,t)}|,
\end{eqnarray}
and
\begin{equation}
\rho_{field}^{out}(t\rightarrow +\infty)\;=\; |1_{\xi}\rangle \langle 1_{\xi}|\,.
\end{equation}

Let us consider now the situation when the ancilla is initially in the state
\begin{equation}\label{ini}
\rho_{A}(0)\;=\; \gamma_{00}|0\rangle\langle 0|+\gamma_{01}|1\rangle \langle 0|+\gamma_{10}|0\rangle \langle 1|+\gamma_{11}|1\rangle \langle 1|\,.
\end{equation}
One can check that in this case the state of the Bose field after interaction with the ancilla till the time $t$ has the form
\begin{eqnarray}
\rho_{field}^{out}(t)&=& \gamma_{00} |\mathit{vac}_{[0,t)}\rangle \langle \mathit{vac}_{[0,t)}|+\gamma_{01} | 1_{\xi[0,t)} \rangle\langle\mathit{vac}_{[0,t)}|\nonumber\\
&&+ \gamma_{10} |\mathit{vac}_{[0,t)}\rangle \langle 1_{\xi[0,t)}|
+\gamma_{11} |1_{\xi[0,t)}\rangle \langle 1_{\xi[0,t)}|\nonumber\\&&+\gamma_{11} \int_{t}^{+\infty}|\xi(s)|^2ds |\mathit{vac}_{[0,t)}\rangle \langle \mathit{vac}_{[0,t)}|\,
\end{eqnarray}
and in the limit of long times, we get
\begin{eqnarray}
\rho_{field}^{out}(t\rightarrow +\infty)=\gamma_{00} |\mathit{vac}\rangle \langle \mathit{vac}|+\gamma_{01} | 1_{\xi} \rangle\langle\mathit{vac}|+ \gamma_{10} |\mathit{vac}\rangle \langle 1_{\xi}|+\gamma_{11}  |1_{\xi}\rangle \langle 1_{\xi}|\,.
\end{eqnarray}

Let us note that the expression (\ref{lambda}) we derived under the assumption that the denominator is different from zero. We assume that $\lambda(t)=0$ whenever $\xi(t)=0$, hence we have also  $\displaystyle{\lim_{t\to+\infty}}\lambda(t)=0$

\subsection{Master equation for the extended system}

We consider a cascaded system consisting of the system $\mathcal{S}$ and the ancilla producing the field in the desired non-classical state. The ancilla system is driven by the Bose field in the vacuum state and the output from the ancilla becomes the input field for  $\mathcal{S}$. In cascaded quantum systems the output of the first system is supplied into the second system, but the reverse process is forbidden. When we omit a time shift due to a traveling between the ancilla and $\mathcal{S}$, we get the master equation for the extended system of the form \cite{GZ10}
\begin{eqnarray}\label{masterextend}
\dot{\tilde{\rho}}(t)&=& \mathcal{L}\tilde{\rho}(t)+\mathcal{L}_{A}\tilde{\rho}(t)+
\left[SL_{A}\tilde{\rho}(t),L^{\dagger}\right]+\left[L, \tilde{\rho}(t)L_{A}^{\dagger}S^{\dagger}\right]\nonumber\\
&&+\left(SL_{A}\tilde{\rho}(t)L_{A}^{\dagger}S^{\dagger}-
L_{A}\tilde{\rho}(t)L_{A}^{\dagger}\right)\,,
\end{eqnarray}
where
\begin{equation}
\mathcal{L}_{A}\tilde{\rho}\;=\; L_{A}\tilde{\rho}L^{\dagger}_{A}-\frac{1}{2}L^{\dagger}_{A}L_{A}\tilde{\rho}-
\frac{1}{2}\tilde{\rho} L^{\dagger}_{A} L_{A}\,.
\end{equation}
We make the assumption that the extended system is initially prepared in the product state
$$\tilde{\rho}(t=0)\;=\; \rho(0) \otimes\rho_{A}(0)\,,$$
where $\rho_{A}(0)$ is the initial state of ancilla given by (\ref{ini}).

Now by taking the partial trace of both sides of (\ref{masterextend}) over the Hilbert space of ancilla, we obtain the equation
\begin{eqnarray}\label{master2}
\dot{\rho_{\mathcal{S}}}(t)&=& \mathcal{L}\rho_{\mathcal{S}}(t)
+\left[S\,\mathrm{Tr}_{\mathfrak{h}_{A}}\left(L_{A}\tilde{\rho}(t)\right),L^{\dagger}\right]
+\left[L, \mathrm{Tr}_{\mathfrak{h}_{A}}\left(\tilde{\rho}(t)L_{A}^{\dagger}\right)S^{\dagger}\right]\nonumber\\
&&+\left(S\,\mathrm{Tr}_{\mathfrak{h}_{A}}\left(L_{A}\tilde{\rho}(t)L_{A}^{\dagger}\right)S^{\dagger}
-\mathrm{Tr}_{\mathfrak{h}_{A}}\left(L_{A}\tilde{\rho}(t)L_{A}^{\dagger}\right)\right)
\end{eqnarray}
for $\rho_{\mathcal{S}}(t)\;=\; \mathrm{Tr}_{\mathfrak{h}_{A}}\tilde{\rho}(t)$, which describes the reduced dynamics of $\mathcal{S}$.

To determine the solution to the above equation, we need to find the operators
$\mathrm{Tr}_{\mathfrak{h}_{A}}\left(L_{A}\tilde{\rho}(t)\right)$, $\mathrm{Tr}_{\mathfrak{h}_{A}}\left(\tilde{\rho}(t)L_{A}^{\dagger}\right)$, and $\mathrm{Tr}_{\mathfrak{h}_{A}}\left(L_{A}\tilde{\rho}(t)L_{A}^{\dagger}\right)$.
For this purpose we derive the differential equations for the operators defined as
\begin{eqnarray}
\rho_{\mathcal{S}}^{-}(t)&=&\frac{1}{\xi(t)}\mathrm{Tr}_{\mathfrak{h}_{A}}\left(L_{A}\tilde{\rho}(t)\right)
\nonumber\\&&=\frac{1}{\sqrt{\int_{t}^{+\infty}|\xi(s)|^2ds}}
\mathrm{Tr}_{\mathfrak{h}_{A}}\left(\sigma_{-}\tilde{\rho}(t)\right)\,,
\end{eqnarray}
\begin{eqnarray}
\rho_{\mathcal{S}}^{+}(t)&=&\frac{1}{\xi(t)}\mathrm{Tr}_{\mathfrak{h}_{A}}
\left(L_{A}^{\dagger}\tilde{\rho}(t)\right)
\nonumber\\&&=\frac{1}{\sqrt{\int_{t}^{+\infty}|\xi(s)|^2ds}}
\mathrm{Tr}_{\mathfrak{h}_{A}}\left(\sigma_{+}\tilde{\rho}(t)\right)\,,
\end{eqnarray}
\begin{eqnarray}
\rho_{\mathcal{S}}^{\mp}(t)&=&\frac{1}{\left|\xi(t)\right|^2}
\mathrm{Tr}_{\mathfrak{h}_{A}}\left(L_{A}\tilde{\rho}(t)L_{A}^{\dagger}\right)
\nonumber\\&&=\frac{1}{\int_{t}^{+\infty}|\xi(s)|^2ds}
\mathrm{Tr}_{\mathfrak{h}_{A}}\left(\sigma_{-}\tilde{\rho}(t)\sigma_{+}\right)\,.
\end{eqnarray}
If $\xi(t)=0$, we put simply $\rho_{\mathcal{S}}^{-}(t)=\rho_{\mathcal{S}}^{+}(t)=\rho_{\mathcal{S}}^{\mp}(t)=0$.
It is not difficult to check that they satisfy the following set of differential equations
\begin{equation}
\dot{\rho}_{\mathcal{S}}^{-}(t)\;=\; \mathcal{L}\rho_{\mathcal{S}}^{-}(t)
+\left[L,\rho_{\mathcal{S}}^{\mp}(t)S^{\dagger}\right]\xi^{\ast}(t)\,,
\end{equation}
\begin{equation}
\dot{\rho}_{\mathcal{S}}^{+}(t)\;=\; \mathcal{L}\rho_{\mathcal{S}}^{+}(t)
+\left[S\rho_{\mathcal{S}}^{\mp}(t),L^{\dagger}\right]\xi(t)\,,
\end{equation}
\begin{equation}
\dot{\rho}_{\mathcal{S}}^{\mp}(t)\;=\; \mathcal{L}\rho_{\mathcal{S}}^{\mp}(t)
\end{equation}
with the initial conditions
${\rho}_{\mathcal{S}}^{-}(0)= \gamma_{01}\rho(0)$, ${\rho}_{\mathcal{S}}^{+}(0)= \gamma_{10}\rho(0)$, and ${\rho}_{\mathcal{S}}^{\mp}(0)= \gamma_{11}\rho(0)$.
We see that $(\rho_{\mathcal{S}}^{-}(t))^{\dagger}=\rho_{\mathcal{S}}^{+}(t)$. Note that for all $t\geq 0$ we have $\mathrm{Tr}_{\mathfrak{h}_{s}}\rho_{\mathcal{S}}(t)=1$, $\mathrm{Tr}_{\mathfrak{h}_{s}}\rho_{\mathcal{S}}^{-}(t)=\gamma_{01}$, $\mathrm{Tr}_{\mathfrak{h}_{s}}\rho_{\mathcal{S}}^{+}(t)=\gamma_{10}$, and $\mathrm{Tr}_{\mathfrak{h}_{s}}\rho_{\mathcal{S}}^{\mp}(t)=\gamma_{11}$.
Thus to specify the reduced state of $\mathcal{S}$, one has to solve the set of four coupled differential equations for $\rho_{\mathcal{S}}(t)$, $\rho_{\mathcal{S}}^{-}(t)$, $\rho_{\mathcal{S}}^{+}(t)$, and $\rho_{\mathcal{S}}^{\mp}(t)$.

The reduced dynamics given by (\ref{master2}) is equivalent to the reduced dynamics described in Part 2.2. To show this one has to write the differential equation for $\rho(t)$ by adding Eqs. (\ref{master1}) with the coefficients: $\gamma_{00}$, $\gamma_{01}$, $\gamma_{10}$, and $\gamma_{11}$. Then one can see that: $\rho_{\mathcal{S}}^{-}(t)=\gamma_{11}\rho^{01}(t)+\gamma_{01}\rho^{00}(t)$, $\rho_{\mathcal{S}}^{+}(t)=\gamma_{11}\rho^{10}(t)+\gamma_{10}\rho^{00}(t)$, ${\rho}_{\mathcal{S}}^{\mp}(t)= \gamma_{11}\rho^{00}(t)$, and $\rho(t)=\rho _{\mathcal{S}}(t)$, which ends the proof.

In our model of cascaded systems, the mean values of photons leaving the system $\mathcal{S}$ in the interval $t$ to $t+dt$ has the form
\begin{eqnarray}
\langle d\Lambda^{out}_{t}\rangle&=&\mathrm{Tr}_{\mathfrak{h}_{\mathcal{S}}}\left(L^{\dagger}L\rho_{\mathcal{S}}(t)\right)dt+
\mathrm{Tr}_{\mathfrak{h}_{\mathcal{S}}}\left(L^{\dagger}S\rho^{-}_{S}(t)\right)\xi(t)dt\nonumber\\
&&+\mathrm{Tr}_{\mathfrak{h}_{\mathcal{S}}}\left(S^{\dagger}L\rho^{+}_{S}(t)\right)\xi^{\ast}(t)dt+
\gamma_{11}|\xi(t)|^2dt
\end{eqnarray}
and for the mean value of the optical quadrature in the interval $t$ to $t+dt$ we get the formula
\begin{eqnarray}
\langle dY(t)\rangle &=&\mathrm{Tr}_{\mathfrak{h}_{\mathcal{S}}}\left(S\rho^{-}_{S}(t)\right)\xi(t)dt+
\mathrm{Tr}_{\mathfrak{h}_{\mathcal{S}}}\left(S^{\dagger}\rho^{+}_{S}(t)\right)\xi^{\ast}(t)dt\nonumber\\
&&+\mathrm{Tr}_{\mathfrak{h}_{\mathcal{S}}}\left((L^{\dagger}+L)\rho_{\mathcal{S}}(t)\right)dt
\end{eqnarray}
Note that both expressions are in agreement with the result derived in Part 2.3.

\subsection{The quantum trajectories for the photon counting}

Quantum filtering equation for a system coupled to the Bose field in the vacuum is well known and we can easily write it down for the extended system consisting of the ancilla and $\mathcal{S}$. The stochastic master equation for the extended system and the direct observation of photons leaving the system $\mathcal{S}$ has the form
\begin{eqnarray}\label{filterext1}
d{\hat{\rho}}(t)&=& \mathcal{L}\hat{\rho}(t)dt+\mathcal{L}_{A}\hat{\rho}(t)dt+[SL_{A}\hat{\rho}(t),L^{\dagger}]dt+[L, \hat{\rho}(t)L_{A}^{\dagger}S^{\dagger}]dt\nonumber\\
&&+\left(SL_{A}\hat{\rho}(t)L_{A}^{\dagger}S^{\dagger}-
L_{A}\hat{\rho}(t)L_{A}^{\dagger}\right)dt\nonumber\\
&&+\bigg\{\frac{(L+SL_{A})\hat{\rho}(t)(L+SL_{A})^{\dagger}
}{k_{t}}-\hat{\rho}(t)\bigg\}dN(t),
\end{eqnarray}
where
\begin{equation}
dN(t)=d\Lambda_{t}^{out}-k_{t}dt,
\end{equation}
and
\begin{equation}
k_{t}=\mathrm{Tr}_{\mathfrak{h}_{\mathcal{S}}\otimes\mathfrak{h}_{A}}[(L+SL_{A})^{\dagger}(L+SL_{A})\hat{\rho}(t)].
\end{equation}

The matrix $\hat{\rho}(t)$ is the {\it a posteriori} state of the extended system at $t$ conditioned by the results of all measurements performed till the time $t$. Here $d\Lambda^{out}_{t}$ is the output process for the cascaded system with the posterior mean value equal to $k_{t}dt$. In other words, the quantity $k_{t}dt$ is the mean value of counts in the interval $t$ to $t+dt$ conditional upon the trajectory (the history of all counts) up to time $t$. Note that $(dN(t))^2=dN(t)$ and the mean $\langle dN(t)\rangle=0$, which follows from the properties of $d\Lambda^{out}_{t}$.

Using the characteristic functional method \cite{Bar06}, we can find now the whole statistics of the output counting process. Let us recall that we deal here with the regular counting process which means that at most one photon can be observed at a period of the length $dt$. The probability of having no counts in the time-interval $(0,t]$ is given by the expression
\begin{equation}\label{prob}
P_{0}^{t}(0)=
\mathrm{Tr}_{\mathfrak{h}_{\mathcal{S}}\otimes\mathfrak{h}_{A}}[\Upsilon(t,0)\rho(0)\otimes\rho_{A}(0)]\,,
\end{equation}
where
$\Upsilon(t,s)$, $t\geq s$, is defined by
\begin{equation}
\frac{d}{dt}\Upsilon(t,s)=\tilde{\mathcal{L}}_{t}\Upsilon(t,s)
\end{equation}
with the condition $\Upsilon(s,s)=1$, and
\begin{equation}
\tilde{\mathcal{L}}_{t}\rho=-iH_{\mathrm{eff}}\rho+i\rho H_{\mathrm{eff}}^{\dagger}\,,
\end{equation}
where $H_{\mathrm{eff}}$ is the effective Hamiltonian having the form
\begin{equation}
H_{\mathrm{eff}}=H_{\mathcal{S}}-\frac{i}{2}(L^{\dagger}L+L_{A}^{\dagger}L_{A}+2L_{A}L^{\dagger}S)\,.
\end{equation}
One can check that $P_{0}^{t=0}(0)=1$. Of course, the quantity $1-P_{0}^{t}(0)$ is the probability of at least one count in the interval $(0,t]$.
The multi-time probability density of a count at time $t_{1}$, a count at time $t_{2}$, ..., $(0<t_{1}<t_{2}<\ldots t_{n}<t)$ and no other counts in the interval from $0$ to $t$ is given by
\begin{eqnarray}
p_{0}^{t}(t_{1};t_{2}, \ldots,t_{n})&=&\mathrm{Tr}_{\mathfrak{h}_{\mathcal{S}}\otimes\mathfrak{h}_{A}}[
\Upsilon(t,t_{n})\mathcal{J}(t_{n})\Upsilon(t_{n},t_{n-1})\ldots
\nonumber\\&&\ldots\Upsilon(t_{2},t_{1})\mathcal{J}(t_{1})\Upsilon(t_{1},0)\rho(0)\otimes\rho_{A}(0)]\,,
\end{eqnarray}
where
\begin{equation}
\mathcal{J}(t_{i})\rho=\left(L+SL_{A}(t_{i})\right)\rho\left(L+SL_{A}(t_{i})\right)^{\dagger}\,.
\end{equation}
Our notation in the above expression indicates the fact that the operator $L_{A}$ depends on time.

Now taking the partial trace of Eq. (\ref{filterext1}) over the Hilbert space of ancilla, we obtain the filtering equation for the system $\mathcal{S}$:
 \begin{eqnarray}\label{filter1}
d{\hat{\rho}}_{\mathcal{S}}(t)&=& \mathcal{L}\hat{\rho}_{\mathcal{S}}(t)dt+[S\hat{\rho}_{\mathcal{S}}^{-}(t),L^{\dagger}]\xi(t)dt
+[L, \hat{\rho}_{\mathcal{S}}^{+}(t)S^{\dagger}]\xi^{\ast}(t)dt\nonumber\\
&&+\left(S\hat{\rho}_{\mathcal{S}}^{\mp}(t)S^{\dagger}- \hat{\rho}_{\mathcal{S}}^{\mp}(t)\right)|\xi(t)|^2dt\nonumber\\
&&+\bigg\{\frac{1}{k_{t}}\left[
L\hat{\rho}_{\mathcal{S}}(t)L^{\dagger}+S\hat{\rho}_{\mathcal{S}}^{-}(t) L^{\dagger}\xi(t)\right.\nonumber\\&&\left.+L\hat{\rho}_{\mathcal{S}}^{+}(t)S^{\dagger}\xi^{\ast}(t)
+S\hat{\rho}_{\mathcal{S}}^{\mp}(t)S^{\dagger}|\xi(t)|^2\right]
-\hat{\rho}_{\mathcal{S}}(t)\bigg\}dN(t),
\end{eqnarray}
where the operators
\begin{equation}
\hat{\rho}_{\mathcal{S}}^{-}(t)\;=\; \frac{1}{\xi(t)}\mathrm{Tr}_{\mathfrak{h}_{A}}\left(L_{A}\hat{\rho}(t)\right)\;=\;
\frac{\mathrm{Tr}_{\mathfrak{h}_{A}}\left(\sigma_{-}\hat{\rho}(t)\right)}{\sqrt{\int_{t}^{+\infty}|\xi(s)|^2ds}}\,,
\end{equation}
\begin{equation}
\hat{\rho}_{\mathcal{S}}^{+}(t)\;=\; \frac{1}{\xi(t)}\mathrm{Tr}_{\mathfrak{h}_{A}}\left(L_{A}\hat{\rho}(t)\right)\;=\;
\frac{\mathrm{Tr}_{\mathfrak{h}_{A}}\left(\sigma_{+}\hat{\rho}(t)\right)}{\sqrt{\int_{t}^{+\infty}|\xi(s)|^2ds}}\,,
\end{equation}
\begin{equation}
\hat{\rho}_{\mathcal{S}}^{\mp}(t)\;=\;\frac{1}{\left|\xi(t)\right|^2}
\mathrm{Tr}_{\mathfrak{h}_{A}}\left(L_{A}\hat{\rho}(t)L_{A}^{\dagger}\right)\;=\;
\frac{\mathrm{Tr}_{\mathfrak{h}_{A}}\left(\sigma_{-}\hat{\rho}(t)\sigma_{+}\right)}{\int_{t}^{+\infty}|\xi(s)|^2ds}
\end{equation}
satisfy the set of the coupled stochastic equations
\begin{eqnarray}
d{\hat{\rho}}_{\mathcal{S}}^{-}(t)&=& \mathcal{L}\hat{\rho}_{\mathcal{S}}^{-}(t)dt
+\left[L,\hat{\rho}_{\mathcal{S}}^{\mp}(t)S^{\dagger}\right]\xi^{\ast}(t)dt\nonumber\\
&&+\left\{\frac{1}{k_{t}}\left(L\hat{\rho}_{\mathcal{S}}^{-}(t)L^{\dagger}+
L\hat{\rho}_{\mathcal{S}}^{\mp}(t)S^{\dagger}
\xi^{\ast}(t)\right)-\hat{\rho}_{\mathcal{S}}^{-}(t)\right\}dN(t),
\end{eqnarray}
\begin{eqnarray}
d{\hat{\rho}}_{\mathcal{S}}^{+}(t)&=& \mathcal{L}\hat{\rho}_{\mathcal{S}}^{+}(t)dt
+\left[S\hat{\rho}_{\mathcal{S}}^{\mp}(t), L^{\dagger}\right]\xi(t)dt\nonumber\\
&&+\left\{\frac{1}{k_{t}}\left(L\hat{\rho}_{\mathcal{S}}^{+}(t)L^{\dagger}+
S\hat{\rho}_{\mathcal{S}}^{\mp}(t)L^{\dagger}
\xi(t)\right)-\hat{\rho}_{\mathcal{S}}^{+}(t)\right\}dN(t),
\end{eqnarray}
\begin{eqnarray}
d{\hat{\rho}}_{\mathcal{S}}^{\mp}(t)&=& \mathcal{L}\hat{\rho}_{\mathcal{S}}^{\mp}(t)dt+\left(
\frac{1}{k_{t}}L\hat{\rho}_{\mathcal{S}}^{\mp}(t)L^{\dagger}-
\hat{\rho}_{\mathcal{S}}^{\mp}(t)\right)dN(t),
\end{eqnarray}
with the initial condition $\hat{{\rho}}_{\mathcal{S}}^{-}(0)= \gamma_{01}\rho(0)$, $\hat{{\rho}}_{\mathcal{S}}^{+}(0)= \gamma_{10}\rho(0)$, $\hat{{\rho}}_{\mathcal{S}}^{\mp}(0)= \gamma_{11}\rho(0)$, and $\hat{\rho}_{\mathcal{S}}(0)=\rho(0)$. One can check easily that the intensity $k_{t}$ is equal to
\begin{eqnarray}
k_{t}=\mathrm{Tr}_{\mathfrak{h}_{\mathcal{S}}}\left[L^{\dagger}L\hat{\rho}_{\mathcal{S}}(t)+
L^{\dagger}S{\hat{\rho}}_{\mathcal{S}}^{-}(t)\xi(t)+S^{\dagger}L{\hat{\rho}}_{\mathcal{S}}^{+}(t)\xi^{\ast}(t)+
|\xi(t)|^2\hat{{\rho}}_{\mathcal{S}}^{\mp}(t)\right].
\end{eqnarray}

Eq. (\ref{prob}) implies that the probability of having no counts in the time interval from $0$ to $t$ can be expressed as $P_{0}^{t}(0)=\mathrm{Tr}_{\mathfrak{h}_{\mathcal{S}}}\hat{\rho}_{\mathcal{S}}(t)$ with  $\hat{\rho}_{\mathcal{S}}(t)$ which can be obtained by solving the system of differential equations
 \begin{eqnarray}\label{prob1}
{\dot{\hat{\rho}}}_{S}(t)&=& -i[H_{\mathcal{S}},\hat{\rho}_{\mathcal{S}}(t)]-\frac{1}{2}L^{\dagger}L\hat{\rho}_{\mathcal{S}}(t)- \frac{1}{2}\hat{\rho}_{\mathcal{S}}(t)L^{\dagger}L\nonumber\\
&&-L^{\dagger}S\hat{\rho}_{\mathcal{S}}^{-}(t)\xi(t)-\hat{\rho}_{\mathcal{S}}^{+}(t)S^{\dagger}L\xi^{\ast}(t)
-\hat{\rho}_{\mathcal{S}}^{\mp}(t)|\xi(t)|^2,
\end{eqnarray}
\begin{eqnarray}\label{prob2}
\dot{\hat{\rho}}_{\mathcal{S}}^{-}(t)&=& -i[H_{\mathcal{S}},\hat{\rho}_{\mathcal{S}}^{-}(t)]-\frac{1}{2}L^{\dagger}L\hat{\rho}_{\mathcal{S}}^{-}(t)
-\frac{1}{2}\hat{\rho}_{\mathcal{S}}^{-}(t)L^{\dagger}L\nonumber\\
&&-\hat{\rho}_{\mathcal{S}}^{\mp}(t)S^{\dagger}L\xi^{\ast}(t),
\end{eqnarray}
\begin{eqnarray}\label{prob3}
\dot{\hat{\rho}}_{\mathcal{S}}^{+}(t)&=& -i[H_{\mathcal{S}},\hat{\rho}_{\mathcal{S}}^{+}(t)]-\frac{1}{2}L^{\dagger}L\hat{\rho}_{\mathcal{S}}^{+}(t)
-\frac{1}{2}\hat{\rho}_{\mathcal{S}}^{+}(t)L^{\dagger}L\nonumber\\
&&-L^{\dagger}S\hat{\rho}_{\mathcal{S}}^{\mp}(t)\xi(t),
\end{eqnarray}
\begin{eqnarray}\label{prob4}
\dot{\hat{\rho}}_{\mathcal{S}}^{\mp}(t)&=&-i[H_{\mathcal{S}},\hat{\rho}_{\mathcal{S}}^{\mp}(t)] -\frac{1}{2}L^{\dagger}L\hat{\rho}_{\mathcal{S}}^{\mp}(t)-
\frac{1}{2}\hat{\rho}_{\mathcal{S}}^{\mp}(t)L^{\dagger}L
\end{eqnarray}
with $\hat{{\rho}}_{\mathcal{S}}^{-}(0)= \gamma_{01}\rho(0)$, $\hat{{\rho}}_{\mathcal{S}}^{+}(0)= \gamma_{10}\rho(0)$, $\hat{{\rho}}_{\mathcal{S}}^{\mp}(0)= \gamma_{11}\rho(0)$,  and $\hat{\rho}_{\mathcal{S}}(0)=\rho(0)$.

We found the conditional evolution of $\mathcal{S}$ depending on the results of measurement performed on the output Bose field. If these results are not read, i.e. no selection is made, the state of the system $\mathcal{S}$ at time $t$ is given by $\rho_{\mathcal{S}}(t)$ which fulfils Eq. (\ref{master2}). We can obtain the master equation for $\mathcal{S}$ by taking the stochastic mean of Eq. (\ref{filter1}).

Note that when $\gamma_{01}=\gamma_{10}=\gamma_{11}=0$ and $\gamma_{00}=1$, we get $\hat{{\rho}}_{\mathcal{S}}^{-}(t)=\hat{{\rho}}_{\mathcal{S}}^{-}(t)=
\hat{{\rho}}_{\mathcal{S}}^{\mp}(t)= 0$ and  for all $t$ and Eq. (\ref{filter1}) reduces then to the filtering equation for the Bose field in the vacuum state. When $\gamma_{01}=\gamma_{10}=\gamma_{00}=0$ and $\gamma_{11}=1$, we obtain the Bose field in a single photon state and our filtering equation reduces to the filtering equation derived in \cite{GJN11,GJNC12}.

\subsection{The quantum trajectories for the quadrature measurement}

The conditional evolution of the extended system for the case of the quadrature measurement of the output field is described by the stochastic equation
\begin{eqnarray}\label{filterext2}
d{\hat{\rho}}(t)&=& \mathcal{L}\hat{\rho}(t)dt+\mathcal{L}_{A}\hat{\rho}(t)dt+[SL_{A}\hat{\rho}(t),L^{\dagger}]dt
+[L, \hat{\rho}(t)L_{A}^{\dagger}S^{\dagger}]dt\nonumber\\
&&+\left(SL_{A}\hat{\rho}(t)L_{A}^{\dagger}S^{\dagger}-
L_{A}\hat{\rho}(t)L_{A}^{\dagger}\right)dt\nonumber\\
&&+\left(\left(L+SL_{A}\right)\hat{\rho}(t)+
\hat{\rho}(t)\left(L+SL_{A}\right)^{\dagger}-
v_{t}\hat{\rho}(t)\right)dW(t),
\end{eqnarray}
where
\begin{equation}
dW(t)= dY(t)-v_{t}dt,
\end{equation}
\begin{equation}
v_{t}=\mathrm{Tr}_{\mathfrak{h}_{\mathcal{S}}\otimes\mathfrak{h}_{A}}
[(L+L^{\dagger}+SL_{A}+L_{A}^{\dagger}S^{\dagger})\hat{\rho}(t)],
\end{equation}
and $dY(t)=dB^{out}_{t}+dB^{out\dagger}_{t}$. Here $dB^{out}_{t}$, $dB^{out\dagger}_{t}$ stand for the output field from the cascaded system.

Equation (\ref{filterext2}) sets the posterior state $\hat{\rho}(t)$ of the extended system at time $t$ depending on all results of the measurement optical quadrature up to time $t$.  Note that the posterior means  $\langle dY(t)\rangle=v_{t}dt$, $\langle dW(t)\rangle = 0$, and $\langle (dW(t))^2\rangle = dt$.

Taking the partial trace over the Hilbert space of ancilla from (\ref{filterext2}) we obtain the stochastic evolution of $\mathcal{S}$. The filtering equation for the conditional state of $\mathcal{S}$, $\hat{\rho}_{\mathcal{S}}(t)=\mathrm{Tr}_{\mathfrak{h}_{A}}\hat{\rho}(t)$, has thus the form
 \begin{eqnarray}\label{filter2}
d{\hat{\rho}}_{\mathcal{S}}(t)&=& \mathcal{L}\hat{\rho}_{\mathcal{S}}(t)dt+[S\hat{\rho}_{\mathcal{S}}^{-}(t),L^{\dagger}]\xi(t)dt
+[L, \hat{\rho}_{\mathcal{S}}^{+}(t)S^{\dagger}]\xi^{\ast}(t)dt\nonumber\\
&&+\left(S\hat{\rho}_{\mathcal{S}}^{\mp}(t)S^{\dagger}- \hat{\rho}_{\mathcal{S}}^{\mp}(t)\right)|\xi(t)|^2dt\nonumber\\
&&+\left(
L\hat{\rho}_{\mathcal{S}}(t)+ \hat{\rho}_{\mathcal{S}}(t)L^{\dagger}+S\hat{\rho}^{-}_{S}(t)\xi(t)\right.
\nonumber\\&&\left.+\hat{\rho}^{+}_{S}(t)S^{\dagger}\xi^{\ast}(t)-
v_{t}\hat{\rho}_{\mathcal{S}}(t)\right)dW(t).
\end{eqnarray}
The matrices $\hat{{\rho}}_{\mathcal{S}}^{-}(t)$,  $\hat{{\rho}}_{\mathcal{S}}^{+}(t)$, $\hat{{\rho}}_{\mathcal{S}}^{\mp}(t)$ satisfy the stochastic equations
\begin{eqnarray}
d{\hat{\rho}}_{\mathcal{S}}^{-}(t)&=& \mathcal{L}\hat{\rho}_{\mathcal{S}}^{-}(t)dt
+\left[L,\hat{\rho}_{\mathcal{S}}^{\mp}(t)S^{\dagger}\right]\xi^{\ast}(t)dt\nonumber\\
&&+\left(L\hat{\rho}_{\mathcal{S}}^{-}(t)+
\hat{\rho}_{\mathcal{S}}^{-}(t)L^{\dagger}\right.\nonumber\\&&\left.
+\hat{\rho}_{\mathcal{S}}^{\mp}(t)S^{\dagger}
\xi^{\ast}(t)-v_{t}\hat{\rho}_{\mathcal{S}}^{-}(t)\right)dW(t)\,,
\end{eqnarray}
\begin{eqnarray}
d{\hat{\rho}}_{\mathcal{S}}^{+}(t)&=& \mathcal{L}\hat{\rho}_{\mathcal{S}}^{+}(t)dt
+\left[S\hat{\rho}_{\mathcal{S}}^{\mp}(t)L\right]\xi(t)dt\nonumber\\
&&+\left(L\hat{\rho}_{\mathcal{S}}^{+}(t)+
\hat{\rho}_{\mathcal{S}}^{+}(t)L^{\dagger}\right.\nonumber\\&&\left.
+S\hat{\rho}_{\mathcal{S}}^{\mp}(t)
\xi(t)-v_{t}\hat{\rho}_{\mathcal{S}}^{+}(t)\right)dW(t)\,,
\end{eqnarray}
\begin{eqnarray}
d{\hat{\rho}}_{\mathcal{S}}^{\mp}(t)&=& \mathcal{L}\hat{\rho}_{\mathcal{S}}^{\mp}(t)dt
+\left(L\hat{\rho}_{\mathcal{S}}^{\mp}(t)+
\hat{\rho}_{\mathcal{S}}^{\mp}(t)L^{\dagger}-v_{t}\hat{\rho}_{\mathcal{S}}^{\mp}(t)\right)dW(t)\,,
\end{eqnarray}
and initially $\hat{{\rho}}_{\mathcal{S}}^{-}(0)= \gamma_{01}\rho(0)$, $\hat{{\rho}}_{\mathcal{S}}^{+}(0)= \gamma_{10}\rho(0)$, $\hat{{\rho}}_{\mathcal{S}}^{\mp}(0)= \gamma_{11}\rho(0)$,  and $\hat{\rho}_{\mathcal{S}}(0)=\rho(0)$. Moreover, one can check that
\begin{equation}
v_{t}=\mathrm{Tr}_{\mathfrak{h}_{\mathcal{S}}}
[(L+L^{\dagger})\hat{\rho}_{\mathcal{S}}(t)
+S\hat{\rho}_{\mathcal{S}}^{-}(t)\xi(t)+\hat{\rho}_{\mathcal{S}}^{+}(t)S^{\dagger}\xi^{\ast}(t)].
\end{equation}

\subsection{Numerical example}

As a simple numerical example we consider $\mathcal{S}$ being a two level atom and we assume that $L=\sqrt{\kappa}\sigma_{-}$, $S=I$, and $H_{\mathcal{S}}=0$ (it means that we work in the interaction picture). We make a calculation for the case when the Bose field is characterized by
\begin{equation}
\xi(t)=\frac{1}{\mathcal{N}}\left(\frac{\Omega^2}{2\pi}\right)^{1/4}\exp\left[-\frac{\Omega^2}{4}\left(t-3\right)^2\right],
\end{equation}
where
\begin{equation}
\mathcal{N}=\left(\frac{\Omega^2}{2\pi}\right)^{1/2}\int_{0}^{+\infty}\exp\left[-\frac{\Omega^2}{4}\left(s-3\right)^2\right]ds.
\end{equation}
To get an optimal excitation for the Gaussian pulse we take $\Omega = 1.46$ \cite{WMSS11}. The quantum filer is given then by the set of eleven coupled equations. Using the formulae (\ref{prob1})-(\ref{prob4}) we are able to plot the curve with the time dependence of the probability of at least one count in the interval $(0,t]$. One can deduce from (\ref{prob1})-(\ref{prob4}) the that the probability of at least one count in the interval from $0$ to $+\infty$ is $1 -\langle 0|\rho(0)|0\rangle (1-\gamma_{11}) $.

\begin{figure}[h]
\begin{center}
\includegraphics[width=5cm]{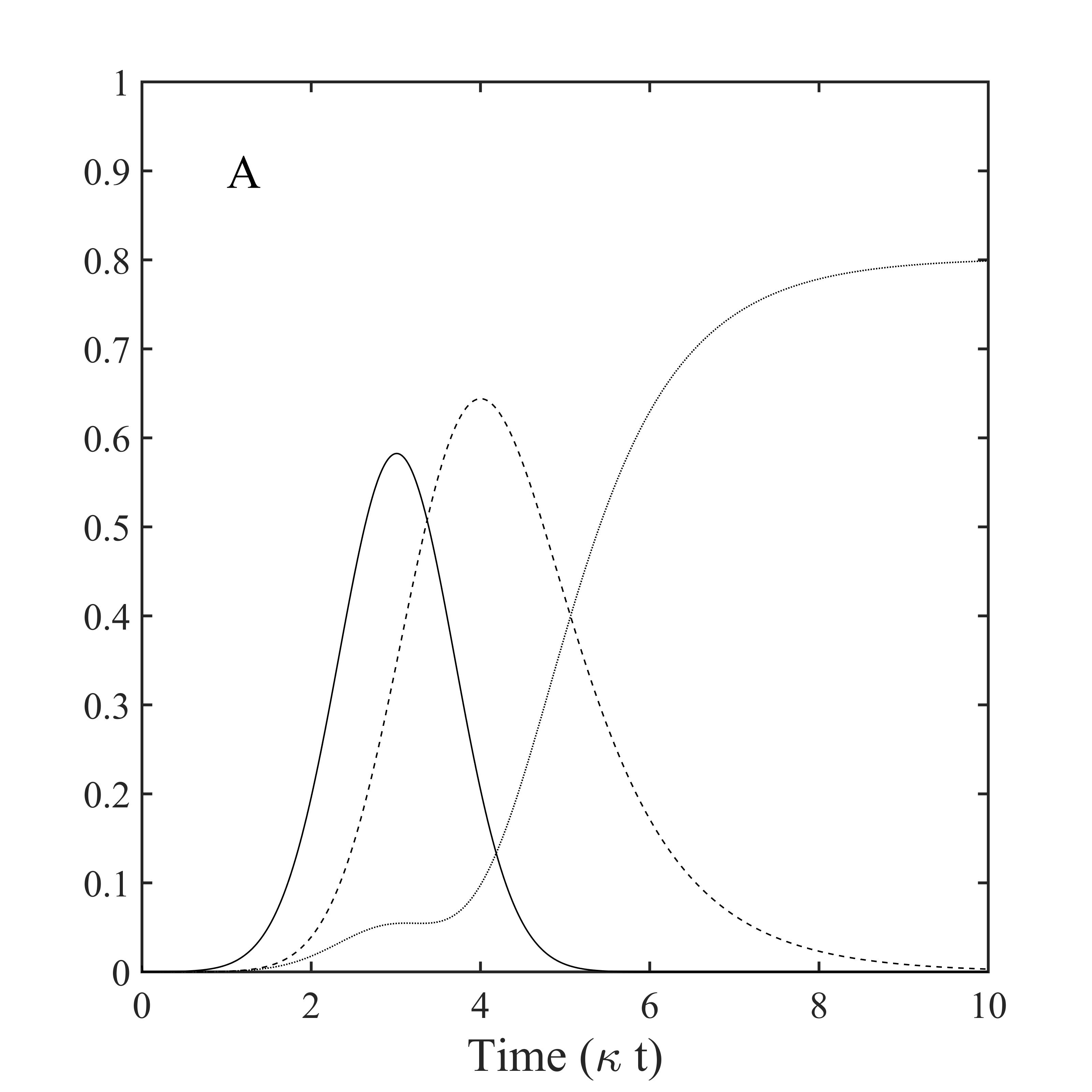}
\includegraphics[width=5cm]{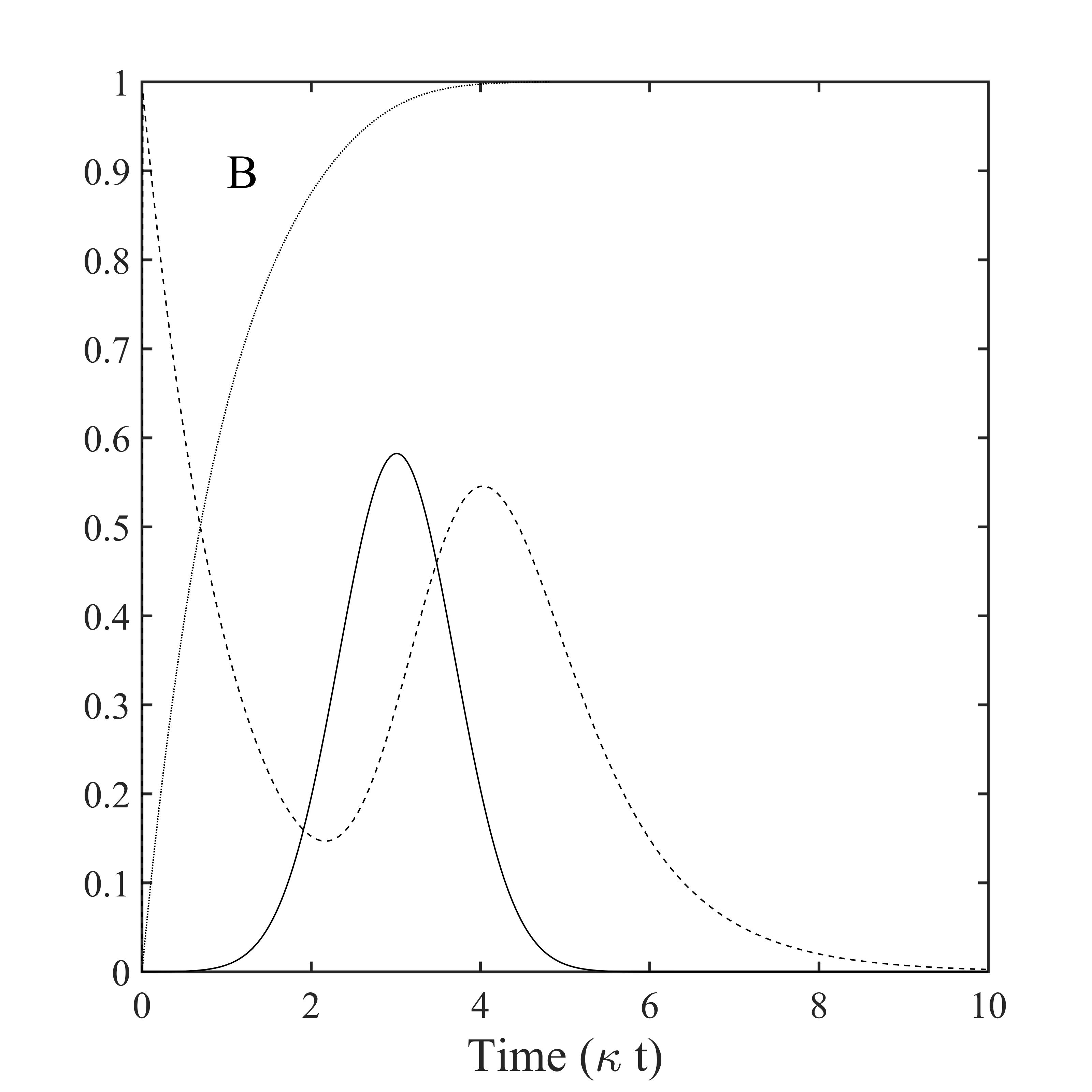}
\caption{The solid line is $|\xi(t)|^2$, the dashed line is the probability of excitation  calculated from the master equation, and the dotted line is the probability of at least one count in the interval $(0,t]$ calculated for the Bose field taken in a mixture of vacuum and single photon states with $\gamma_{11}=0.8$, $\gamma_{00}=0.2$, $\gamma_{01}=\gamma_{10}=0$, and for $\mathcal{S}$ being initially in the ground state (A) and in the upper state (B).}
\end{center}
\end{figure}

\section{Filtering equation for system driven by the Bose field in a mixture of coherent states}

\subsection{Continuous-mode coherent states}

Continuous-mode coherent state is defined as
\begin{equation}
|\alpha\rangle \;=\overrightarrow{\mathrm{T}}\;\exp\left\{\int_{0}^{+\infty}\alpha(t)dB_{t}^{\dagger}-\alpha^{\ast }(t)dB_{t}\right\}|\mathit{vac}\rangle\,,
\end{equation}
where $\overrightarrow{\mathrm{T}}$ stands for the chronological ordering operator. Note that
\begin{eqnarray}
\overrightarrow{\mathrm{T}}\exp\left\{\int_{0}^{t}\alpha(s)dB_{s}^{\dagger}-\alpha^{\ast }(s)dB_{s}\right\}|\mathit{vac}\rangle= \\= |\alpha_{[0,t)}\rangle \otimes |\mathit{vac}_{[t,+\infty)}\rangle\,,
\end{eqnarray}
so that the continuous-mode coherent state has the factorization property
\begin{equation}
|\alpha\rangle\;=\; |\alpha_{[0,t)}\rangle \otimes|\alpha_{[t,+\infty)}\rangle\,.
\end{equation}

The mean values of increments $dB_{t}$, $dB_{t}^{\dagger}$, $d\Lambda_{t}$ for the coherent state are
\begin{equation}
\langle \alpha| dB_{t}|\alpha\rangle = \alpha(t)dt\,,\;\;\;\langle \alpha| d\Lambda_{t}|\alpha\rangle = |\alpha(t)|^2dt,
\end{equation}
$$\langle \alpha|dB_{t}dB^{\dagger}_{t}|\alpha\rangle = dt,\;\;
\langle \alpha|dB^{\dagger}_{t}dB_{t}|\alpha\rangle = 0, \;\;\;
\langle \alpha|d\Lambda_{t}dB^{\dagger}_{t}|\alpha\rangle = \alpha^{\ast}(t)dt.$$

\subsection{Reduced dynamics of $\mathcal{S}$}

One can check that the reduced evolution of an open quantum system interacting with the Bose field in the state
\begin{equation}
\rho_{field}= p|\alpha_{0}\rangle\langle \alpha_{0}|+(1-p)|\alpha_{1}\rangle\langle \alpha_{1}|
\end{equation}
with $p\in [0,1]$, can be written as
\begin{equation}
\rho(t)=p\rho^{00}(t)+(1-p)\rho^{11}(t),
\end{equation}
where the matrices $\rho^{00}(t)$ and $\rho^{11}(t)$ satisfy the following set of differential equations
\begin{eqnarray}
\dot{\rho}^{00}(t)&=& \mathcal{L}\rho^{00}(t)
+\alpha_{0}(t)\left[S\rho^{00}(t),L^{\dagger}\right]+\alpha_{0}^{\ast}(t)\left[L,\rho^{00}(t)S^{\dagger}\right]
\nonumber\\&&+|\alpha_{0}(t)|^2\left(S\rho^{00}(t)S^{\dagger}-\rho^{00}(t)\right),
\end{eqnarray}
\begin{eqnarray}
\dot{\rho}^{11}(t)&=& \mathcal{L}\rho^{11}(t)+
\alpha_{1}(t)\left[S{\rho}^{11}(t),L^{\dagger}\right]
+\alpha_{1}^{\ast}(t)\left[L,{\rho}^{11}(t)S^{\dagger}\right]\nonumber\\
&&+|\alpha_{1}(t)|^2\left(S\rho^{11}(t)S^{\dagger}-\rho^{11}(t)\right)
\end{eqnarray}
with the initial conditions $\rho^{00}(0)\;=\;\rho^{11}(0)\;=\;\rho(0)$ and the superoperator $\mathcal{L}$ given by (\ref{superop}).

\subsection{Generator of the Bose field in a mixture of two coherent states}

Let us consider a two level system interacting with the Bose field in the vacuum state. The unitary evolution of the ancilla and the Bose field is given by Eq. (\ref{uni1}), but this time we assume that the coupling operator is defined as
\begin{equation}\label{coupling2}
L_{A}\;=\; \alpha_{0}(t)|0\rangle \langle 0|+\alpha_{1}(t)|1\rangle \langle 1 |\,,
\end{equation}
where $|0\rangle$, $|1\rangle$ is the orthonormal basis of ancilla.

So that if the initial state of the compound system is
\begin{equation}
\left(c_{0}|0 \rangle+c_{1}|1\rangle\right)\otimes |\mathit{vac}\rangle\,,
\end{equation}
then the system evolves according to the formula
\begin{eqnarray}
|\psi_{t}\rangle&=& c_{0}|0\rangle \otimes |\alpha_{ 0[0,t)}\rangle\otimes|\mathit{vac}_{[t,+\infty)}\rangle\nonumber\\
&&+c_{1}|1\rangle \otimes |\alpha_{1[0,t)}\rangle\otimes|\mathit{vac}_{[t,+\infty)}\rangle
\end{eqnarray}
and in the limit we have
\begin{equation}
\lim_{t\rightarrow +\infty} |\psi_{t}\rangle\;=\;  c_{0}|0\rangle \otimes |\alpha_{0}\rangle+c_{1}|1\rangle \otimes |\alpha_{1}\rangle\,.
\end{equation}
The state of the Bose field after interaction with ancilla till time $t$
has the form
\begin{eqnarray}
\rho_{field}^{out}(t)&=& \left|c_{0}\right|^2 |\alpha_{0[0,t)}\rangle \langle \alpha_{0 [0,t)}|+\left|c_{1}\right|^2 |\alpha_{1[0,t)}\rangle \langle \alpha_{1 [0,t)}|\,.
\end{eqnarray}
This shows that choosing the initial state of the ancilla as
\begin{equation}\label{ini2}
\rho_{A}(0)\;=\; |c_{0}|^2|0\rangle\langle 0|+c_{0}c_{1}^{\ast}|0\rangle\langle 1|+c_{0}^{\ast}c_{1}|1\rangle\langle 0|+|c_{1}|^2|1\rangle \langle 1|
\end{equation}
and the coupling operator (\ref{coupling2}), we obtain the output field in a mixture of two coherent states.

As before, we consider a cascaded system consisting of ancilla and some quantum system $\mathcal{S}$. We obtain dynamics of $\mathcal{S}$ from Eq. (\ref{masterextend}) taking (\ref{coupling2}) and assuming that initially the ancilla is prepared in state (\ref{ini2}). It is not difficult to check that in this case, the reduce dynamics of $\mathcal{S}$ is given by
\begin{equation}
\rho_{\mathcal{S}}(t)\;=\; {\rho}_{\mathcal{S}}^{00}(t)+{\rho}_{\mathcal{S}}^{11}(t)\,,
\end{equation}
where
\begin{equation}
{\rho}_{\mathcal{S}}^{00}(t)=\langle 0| \tilde{\rho}(t)| 0\rangle,\;\;\;
{\rho}_{\mathcal{S}}^{11}(t)=\langle 1| \tilde{\rho}(t)|1\rangle
\end{equation}
satisfy the differential equations
\begin{eqnarray}
\dot{\rho}_{\mathcal{S}}^{00}(t)&=& \mathcal{L}\rho_{\mathcal{S}}^{00}(t)
+\alpha_{0}(t)\left[\rho_{\mathcal{S}}^{00}(t),L^{\dagger}\right]
+\alpha_{0}^{\ast}(t)\left[L,\rho_{\mathcal{S}}^{00}(t)\right]\nonumber\\
&&+|\alpha_{0}(t)|^2\left(S\rho^{00}_{S}(t)S^{\dagger}-\rho^{00}_{S}(t)\right)\,,
\end{eqnarray}
\begin{eqnarray}
\dot{\rho}_{\mathcal{S}}^{11}(t)&=& \mathcal{L}_{S}\rho_{\mathcal{S}}^{11}(t)+
\alpha_{1}(t)\left[{\rho}_{\mathcal{S}}^{11}(t),L^{\dagger}_{S}\right]
+\alpha_{1}^{\ast}(t)\left[L_{S},{\rho}_{\mathcal{S}}^{11}(t)\right]\nonumber\\
&&+|\alpha_{1}(t)|^2\left(S\rho^{11}_{S}(t)S^{\dagger}-\rho^{11}_{S}(t)\right)
\end{eqnarray}
and ${\rho}_{\mathcal{S}}^{00}(0)= |c_{0}|^2\rho_{\mathcal{S}}(0)$, ${\rho}_{\mathcal{S}}^{11}(0)= |c_{1}|^2\rho_{\mathcal{S}}(0)$.

It is seen that when $|c_{0}|^2=p$ and $|c_{1}|^2=1-p$, the above procedure gives the same reduced dynamics as was considered in Part 3.2.

\subsection{The quantum trajectories}

The conditional state of $\mathcal{S}$ may now be expressed by
\begin{equation}
\hat{\rho}_{\mathcal{S}}(t)=\hat{\rho}_{\mathcal{S}}^{00}(t)+\hat{\rho}_{\mathcal{S}}^{11}(t),
\end{equation}
where the matrices
\begin{equation}
\hat{\rho}_{\mathcal{S}}^{00}(t)=\langle 0|\hat{\rho}(t)|0\rangle,\;\;\;
\hat{\rho}_{\mathcal{S}}^{11}(t)=\langle 1|\hat{\rho}(t)|1\rangle
\end{equation}
and $\hat{\rho}(t)$ is the conditional state of the extended system satisfying
(\ref{filterext1}) or (\ref{filterext2}) (it depends on the measurement scheme)
with (\ref{coupling2}) and the initial state of ancilla given as (\ref{ini2}). Thus initially we have $\hat{\rho}_{\mathcal{S}}^{00}(0)= |c_{0}|^2\rho(0)$, $\hat{\rho}_{\mathcal{S}}^{11}(0)=|c_{1}|^2\rho(0)$.

One finds easily that for the photon counting process, the matrices $\hat{\rho}_{\mathcal{S}}^{00}(t)$, $\hat{\rho}_{\mathcal{S}}^{11}(t)$ satisfy the set of stochastic equations
\begin{eqnarray}
d\hat{\rho}_{\mathcal{S}}^{00}(t)&=& \mathcal{L}\hat{\rho}_{\mathcal{S}}^{00}(t)dt
+\alpha_{0}(t)\left[S\hat{\rho}_{\mathcal{S}}^{11}(t),L^{\dagger}\right]dt
+\alpha_{0}^{\ast}(t)\left[L,\hat{\rho}_{\mathcal{S}}^{00}(t)S^{\dagger}\right]dt\nonumber\\
&&+|\alpha_{0}(t)|^2\left(S\hat{\rho}^{00}_{S}(t)S^{\dagger}-\hat{\rho}^{00}_{S}(t)\right)dt\nonumber\\
&&+\left\{\frac{1}{k_{t}}\left(L\hat{\rho}_{\mathcal{S}}^{00}(t)L^{\dagger}+\alpha_{0}^{\ast}(t)L\hat{\rho}_{\mathcal{S}}^{00}(t)S^{\dagger}
+\alpha_{0}(t)S\hat{\rho}_{\mathcal{S}}^{00}(t)L^{\dagger}\right.\right.\nonumber\\
&&\left.\left.+|\alpha_{0}(t)|^2
S\hat{\rho}_{\mathcal{S}}^{00}(t)S^{\dagger}\right)
-\hat{\rho}_{\mathcal{S}}^{00}(t)\right\}dN(t),
\end{eqnarray}
\begin{eqnarray}
d\hat{\rho}_{\mathcal{S}}^{11}(t)&=& \mathcal{L}\hat{\rho}_{\mathcal{S}}^{11}(t)dt+
\alpha_{1}(t)\left[S{\hat{\rho}}_{\mathcal{S}}^{11}(t),L^{\dagger}\right]dt
+\alpha_{1}^{\ast}(t)\left[L,{\hat{\rho}}_{\mathcal{S}}^{11}(t)S^{\dagger}\right]dt\nonumber\\
&&+|\alpha_{1}(t)|^2\left(S\hat{\rho}^{11}_{S}(t)S^{\dagger}-\hat{\rho}^{11}_{S}(t)\right)dt\nonumber\\
&&+\left\{\frac{1}{k_{t}}\left(L\hat{\rho}_{\mathcal{S}}^{11}(t)L^{\dagger}+\alpha_{1}^{\ast}(t)L\hat{\rho}_{\mathcal{S}}^{11}(t)S^{\dagger}
+\alpha_{1}(t)S\hat{\rho}_{\mathcal{S}}^{11}(t)L^{\dagger}\right.\right.\nonumber\\
&&\left.\left.+|\alpha_{1}(t)|^2
S\hat{\rho}_{\mathcal{S}}^{11}(t)S^{\dagger}\right)
-\hat{\rho}_{\mathcal{S}}^{11}(t)\right\}dN(t)\,.
\end{eqnarray}
Here $dN(t)=d\Lambda^{out}_{t}-k_{t}dt$, where $d\Lambda^{out}_{t}$ describes the output number of photons for the cascaded system in the interval from $t$ to $t+dt$ with the posterior mean value $\langle d\Lambda^{out}_{t}\rangle =k_{t}dt$, where
\begin{eqnarray}
k_{t}&=&
\mathrm{Tr}_{\mathfrak{h}_{\mathcal{S}}}\bigg(L^{\dagger}L\hat{\rho}_{\mathcal{S}}(t)
+L^{\dagger}S(\alpha_{0}(t)\hat{\rho}_{\mathcal{S}}^{00}(t)+\alpha_{1}(t)\hat{\rho}_{\mathcal{S}}^{11}(t))
\nonumber\\
&&+S^{\dagger}L(\alpha_{0}^{\ast}(t)\hat{\rho}_{\mathcal{S}}^{00}(t)+\alpha_{1}^{\ast}(t)
\hat{\rho}_{\mathcal{S}}^{11}(t))\nonumber\\
&&+|\alpha_{0}(t)|^2\hat{\rho}_{\mathcal{S}}^{00}(t)+
|\alpha_{1}(t)|^2\hat{\rho}_{\mathcal{S}}^{11}(t)\bigg).
\end{eqnarray}

The probability of having at least one count in the interval $0$ to $t$ is given in this case by $P^{t}_{0}(0)=\mathrm{Tr}_{\mathfrak{h}_{\mathcal{S}}}\left(\hat{\rho}^{00}_{S}(t)
+\hat{\rho}^{00}_{S}(t)\right)$, where
\begin{eqnarray}
\dot{\hat{\rho}}_{\mathcal{S}}^{00}(t)&=& -i[H_{\mathcal{S}},\hat{\rho}_{\mathcal{S}}^{00}(t)]-\frac{1}{2}L^{\dagger}L\hat{\rho}_{\mathcal{S}}^{00}(t)- \frac{1}{2}\hat{\rho}_{\mathcal{S}}^{00}(t)L^{\dagger}L\nonumber\\
&&-L^{\dagger}S\hat{\rho}_{\mathcal{S}}^{00}(t)\alpha_{0}(t)-\hat{\rho}_{\mathcal{S}}^{00}(t)S^{\dagger}L\alpha^{\ast}_{0}(t)\nonumber\\
&&-\hat{\rho}_{\mathcal{S}}^{00}(t)|\alpha_{0}(t)|^2\,,
\end{eqnarray}
\begin{eqnarray}
\dot{\hat{\rho}}_{\mathcal{S}}^{11}(t)&=& -i[H_{\mathcal{S}},\hat{\rho}_{\mathcal{S}}^{11}(t)]-\frac{1}{2}L^{\dagger}L\hat{\rho}_{\mathcal{S}}^{11}(t)- \frac{1}{2}\hat{\rho}_{\mathcal{S}}^{11}(t)L^{\dagger}L\nonumber\\
&&-L^{\dagger}S\hat{\rho}_{\mathcal{S}}^{11}(t)\alpha_{1}(t)\nonumber\\
&&-\hat{\rho}_{\mathcal{S}}^{11}(t)S^{\dagger}L\alpha^{\ast}_{1}(t)
-\hat{\rho}_{\mathcal{S}}^{11}(t)|\alpha_{1}(t)|^2,
\end{eqnarray}
and $\hat{\rho}_{\mathcal{S}}^{00}(0)= |c_{0}|^2\rho(0)$, $\hat{\rho}_{\mathcal{S}}^{11}(0)=|c_{1}|^2\rho(0)$.

One can check that for the evolution of $\mathcal{S}$ conditioned on the results of the measurement of the optical quadrature, the matrices $\hat{\rho}_{\mathcal{S}}^{00}(t)$, $\hat{\rho}_{\mathcal{S}}^{11}(t)$ satisfy the set of stochastic equations
\begin{eqnarray}
d\hat{\rho}_{\mathcal{S}}^{00}(t)&=& \mathcal{L}\hat{\rho}_{\mathcal{S}}^{00}(t)dt
+\alpha_{0}(t)\left[S\hat{\rho}_{\mathcal{S}}^{11}(t),L^{\dagger}\right]dt
+\alpha_{0}^{\ast}(t)\left[L,\hat{\rho}_{\mathcal{S}}^{00}(t)S^{\dagger}\right]dt\nonumber\\
&&+|\alpha_{0}(t)|^2\left(S\hat{\rho}^{00}_{S}(t)S^{\dagger}-\hat{\rho}^{00}_{S}(t)\right)dt\nonumber\\
&&+\left\{L\hat{\rho}_{\mathcal{S}}^{00}(t)+
\hat{\rho}_{\mathcal{S}}^{00}(t)L^{\dagger}+\alpha_{0}(t)S\hat{\rho}_{\mathcal{S}}^{00}(t)\right.
\nonumber\\&&\left.
+\alpha_{0}^{\ast}(t)\hat{\rho}_{\mathcal{S}}^{00}(t)S^{\dagger}-v_{t}\hat{\rho}_{\mathcal{S}}^{00}(t)\right\}dW(t),
\end{eqnarray}
\begin{eqnarray}
d\hat{\rho}_{\mathcal{S}}^{11}(t)&=& \mathcal{L}\hat{\rho}_{\mathcal{S}}^{11}(t)dt+
\alpha_{1}(t)\left[S{\hat{\rho}}_{\mathcal{S}}^{11}(t),L^{\dagger}\right]dt
+\alpha_{1}^{\ast}(t)\left[L,{\hat{\rho}}_{\mathcal{S}}^{11}(t)S^{\dagger}\right]dt\nonumber\\
&&+|\alpha_{1}(t)|^2\left(S\hat{\rho}^{11}_{S}(t)S^{\dagger}-\hat{\rho}^{11}_{S}(t)\right)dt\nonumber\\
&&+\left\{L\hat{\rho}_{\mathcal{S}}^{11}(t)+
\hat{\rho}_{\mathcal{S}}^{11}(t)L^{\dagger}+\alpha_{1}(t)S\hat{\rho}_{\mathcal{S}}^{11}(t)\right.\nonumber\\&&\left.
+\alpha_{1}^{\ast}(t)\hat{\rho}_{\mathcal{S}}^{11}(t)S^{\dagger}-v_{t}\hat{\rho}_{\mathcal{S}}^{11}(t)\right\}dW(t)
\end{eqnarray}
with the initial conditions $\hat{\rho}_{\mathcal{S}}^{00}(0)=|c_{0}|^2\rho_{\mathcal{S}}(0)$,
$\hat{{\rho}}_{\mathcal{S}}^{11}(0)= |c_{1}|^2\rho_{\mathcal{S}}(0)$ and the posterior intensity
\begin{eqnarray}
v_{t}&=&\mathrm{Tr}_{\mathfrak{h}_{\mathcal{S}}}\left((L+L^{\dagger})\hat{\rho}(t)\right)
+\mathrm{Tr}_{\mathfrak{h}_{\mathcal{S}}}\left(S(\alpha_{0}(t)\hat{\rho}^{00}_{S}(t)
+\alpha_{1}(t)\hat{\rho}^{11}_{S}(t))\right)\nonumber\\
&&+\mathrm{Tr}_{\mathfrak{h}_{\mathcal{S}}}\left(S^{\dagger}(\alpha_{0}^{\ast}(t)\hat{\rho}^{00}_{S}(t)
+\alpha_{1}^{\ast}(t)\hat{\rho}^{11}_{S}(t))\right).
\end{eqnarray}
Here $dW=dY-v_{t}dt$ and $\langle dY\rangle =v_{t}dt$.

\subsection{Numerical Example}

Let us consider the coherent Gaussian pulses
\begin{equation}
\alpha_{0}(t) \;=\; \left(\frac{2\Omega^2}{\pi}\right)^{1/4}\exp\left[-\frac{\Omega^2}{4}\left(t-3\right)^2\right]\,,
\end{equation}
\begin{equation}
\alpha_{1}(t) \;=\; \left(\frac{2\Omega^2}{\pi}\right)^{1/4}\exp\left[-\frac{\Omega^2}{4}\left(t-5\right)^2\right]\,.
\end{equation}
with the parameters $\Omega = 2.4\kappa$ and $\kappa=1$. We consider $\mathcal{S}$ being a two level atom and we take $L=\sqrt{\kappa}\sigma_{-}$, $S=I$, and $H_{\mathcal{S}}=0$. In Fig. 2 the probability of being in the excited state as well as the probability of at least one count in the interval $(0,t]$ are shown for two choices of initial conditions. One can check that in the limit we have
\begin{eqnarray}
\lim_{t\to +\infty}P^{t}_{0}(0)=\langle 0|\rho(0)|0\rangle\left[pe^{-\int_{0}^{+\infty}|\alpha_{0}(t)|^2dt}
+(1-p)e^{-\int_{0}^{+\infty}|\alpha_{1}(t)|^2dt}\right].
\end{eqnarray}

\begin{figure}[h]
\begin{center}
\includegraphics[width=5cm]{BB1.jpg}
\includegraphics[width=5cm]{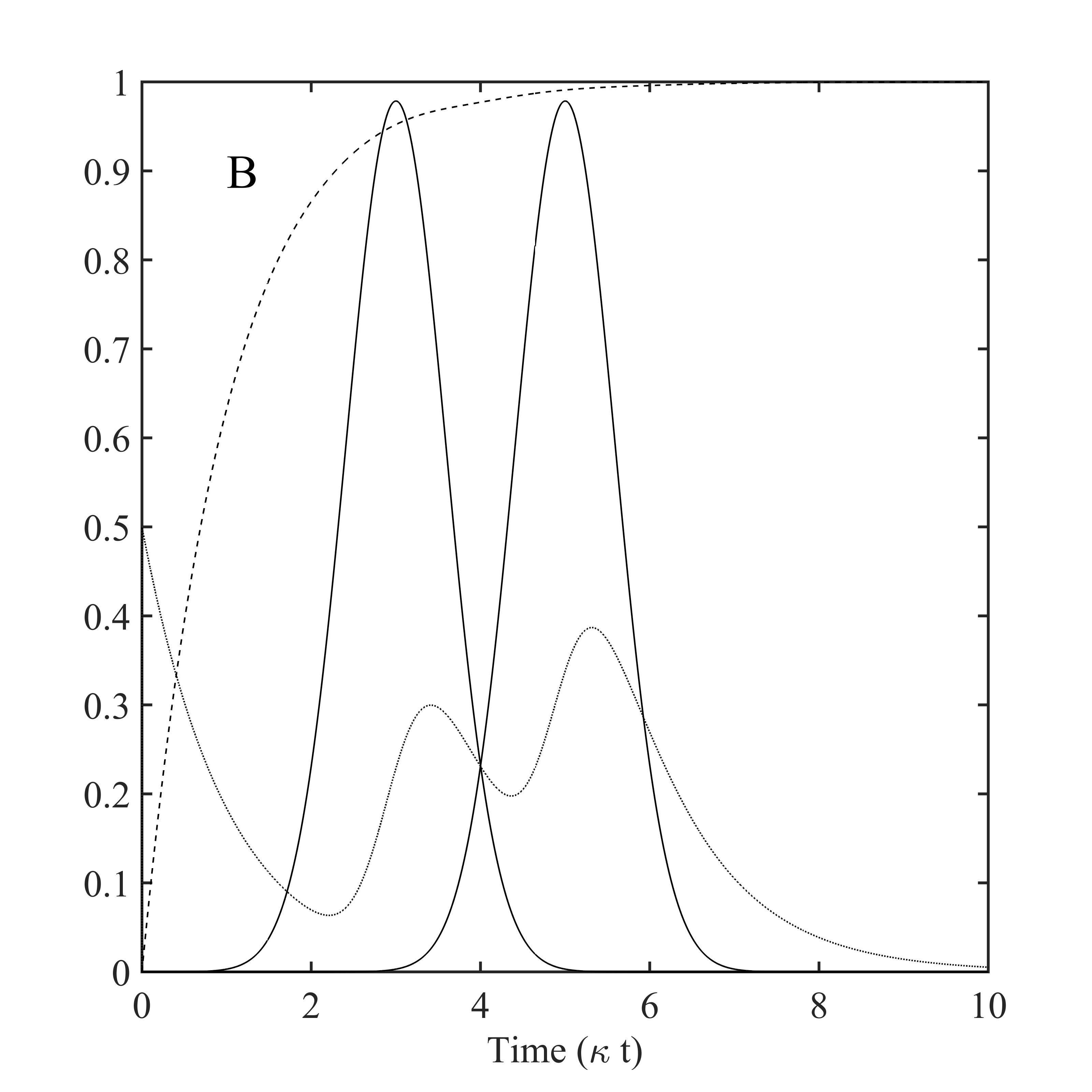}
\caption{The solid lines are $|\alpha_{1}(t)|^2/2$ and $|\alpha_{2}(t)|^2/2$, the dashed line is the probability of excitation from the master equation, and the dotted line is the probability of at least one count in the interval $(0,t]$ for the Bose field taken in a mixture of two coherent states with $p=0.5$ and for $\mathcal{S}$ being initially
in the ground state (A) and in the upper state (B).}
\end{center}
\end{figure}

\section{Conclusions}

We have derived the filtering equations for the two types of non-classical states of the Bose field, namely for a combination of the vacuum and single photon states and for a mixture of two coherent states. To determine the stochastic conditional evolution of an open quantum system we have extended the Hilbert space of this system by ancilla system playing a role of a generator of the Bose field in the desired non-classical state. We have considered the counting as well as the diffusion processes connected respectively with the measurements of the number of photons and the optical quadrature of the output field. For the Bose field taken in a single photon state one can check easily that our filtering equations  agree with the filtering equations derived in \cite{GJNC12}, but for the Bose field taken in a combination of the vacuum and single photon states the proof of the equivalence of the both results requires more effort. At first glance they seem to be inconsistent. Unlike us, the authors of the mentioned paper derived the stochastic evolution for a combination of the vacuum and single photon states assuming that the two-level ancilla system is initially prepared not in a combination of the ground and excited states but in the excited state and to determine the conditional state of $\mathcal{S}$ they used the method of ``weighting and normalisation'' \cite{B04}. Consequences of the different choices of the initial state of ancilla are different intensities of the output processes determining the filters. To use in a proper way the stochastic equations from \cite{GJNC12}, one has to reevaluate the posterior measure for the output processes. In order to prove that our results are consistent with the results presented in \cite{GJNC12} one has to derived, using the standard method of stochastic calculus, the stochastic equations for the matrix $\rho(t)=\sum_{jk}\gamma_{kj}\rho^{jk}(t)/\sum_{jk}\gamma_{kj}\mathrm{Tr}\{\rho^{jk}(t)\}$
from \cite{GJNC12}. They agree with Eq. (\ref{filter1}) and Eq. (\ref{filter2}) (depending on the chosen type of measurement). Of course, our calculations for a mixture of two coherent states can be easily generalised to the case of a mixture of any number of coherent states. Our studies are consistent with the filter given in \cite{GJNC12}, but our procedure can not be used for the Bose field in a superposition of coherent states because the ancilla we considered is not a source of such signal. A proposal of a generator of a superposition of coherent states as well as a description of ancilla generating the Bose field in a multi-photon state one can find in \cite{GZ15}. They can be applied for derivation of the filtering equations for these two non-classical states.

Finally, we would like to mention that there exists more direct, based on the collision model \cite{AP06,PP09,C17}, method for determination of the filtering equations for the quantum noise in non-classical state which does not refer to the concept of ancilla generator. The method was already successfully applied for a single photon state \cite{DSC17} and for a superposition of the coherent states \cite{D17}.

\section{Acknowledgements}
This paper was partially supported by the National
Science Center project 2015/17/B/ST2/02026.

\end{document}